\documentclass[aps,prl,twocolumn,showpacs,superscriptaddress]{revtex4-1}
\usepackage{graphicx}
\usepackage{amsmath}
\usepackage{amssymb}
\usepackage{colordvi}
\usepackage{mathrsfs}
\usepackage{bm}
\usepackage{verbatim}
\usepackage{dcolumn}
\usepackage{bm}
\usepackage{epsfig}
\usepackage{subfigure}
\usepackage{dsfont}
\usepackage[colorlinks,linkcolor=blue,anchorcolor=blue,citecolor=blue]{hyperref}

\begin{document}
\title{Interlayer Ferromagnetism and High-Temperature Quantum Anomalous Hall Effect in \textit{p}-Doped MnBi$_2$Te$_4$ Multilayers}

\author{Yulei Han}
\affiliation{ICQD, Hefei National Laboratory for Physical Sciences at Microscale, CAS Key Laboratory of Strongly-Coupled Quantum Matter Physics, and Department of Physics, University of Science and Technology of China, Hefei, Anhui 230026, China}
\author{Shiyang Sun}
\affiliation{College of Physics, Hebei Normal University, Shijiazhuang, Hebei 050024, China}
\author{Shifei Qi}
\email[Correspondence author:~~]{qisf@hebtu.edu.cn}
\affiliation{College of Physics, Hebei Normal University, Shijiazhuang, Hebei 050024, China}
\affiliation{ICQD, Hefei National Laboratory for Physical Sciences at Microscale, CAS Key Laboratory of Strongly-Coupled Quantum Matter Physics, and Department of Physics, University of Science and Technology of China, Hefei, Anhui 230026, China}
\author{Xiaohong Xu}
\email[Correspondence author:~~]{xuxh@dns.sxnu.edu.cn}
\affiliation{Research Institute of Materials
Science, and School of Chemistry and Materials Science, Shanxi Normal University, Linfen, Shanxi 041004, China}
\author{Zhenhua Qiao}
\email[Correspondence author:~~]{qiao@ustc.edu.cn}
\affiliation{ICQD, Hefei National Laboratory for Physical Sciences at Microscale, CAS Key Laboratory of Strongly-Coupled Quantum Matter Physics, and Department of Physics, University of Science and Technology of China, Hefei, Anhui 230026, China}
\date{\today{}}

\begin{abstract}
  The interlayer antiferromagnetic coupling hinders the observation of quantum anomalous Hall effect in magnetic topological insulator MnBi$_2$Te$_4$. We demonstrate that interlayer \textit{ferromagnetism} can be established by utilizing the \textit{p}-doping method in MnBi$_2$Te$_4$ multilayers. In two septuple-layers system, the interlayer ferromagnetic coupling appears by doping nonmagnetic elements (e.g., N, P, As, Na, Mg, K, and Ca), due to the redistribution of orbital occupations of Mn. We further find that Mg and Ca elements are the most suitable candidates because of their low formation energy. Although, the \textit{p}-doped two septuple layers exhibit topologically trivial band structure, the increase of layer thickness to three (four) septuple layers with Ca (Mg) dopants leads to the formation of the quantum anomalous Hall effect. Our proposed \textit{p}-doping strategy not only makes MnBi$_2$Te$_4$ become an ideal platform to realize the high-temperature quantum anomalous Hall effect without external magnetic field, but also can compensate the electrons from the intrinsic \textit{n}-type defects in MnBi$_2$Te$_4$.
\end{abstract}

\maketitle
\textit{Introduction---.} Quantum anomalous Hall effect (QAHE) is a typical topological quantum phenomena with quantized Hall resistance and vanishing longitudinal resistance in the absence of external magnetic field~\cite{weng2015quantum,ren2016topological,he2018topological}. It is promising in designing low-power electronic devices due to its dissipationless electronic transport properties. Although it was first theoretically proposed by Haldane in 1988~\cite{Haldane}, the exploration of the QAHE began to attract huge interest ever since the first exfoliation of monolayer graphene in 2004~\cite{Graphene2004}. After that, there have been various proposed recipes in designing the QAHE~\cite{proposal1,proposal2,proposal3,proposal4,proposal5,proposal6,proposal7,proposal8,proposal9,proposal10,proposal11,proposal12}, among which the magnetic topological insulator is the most favorable system by both theoretical and experimental studies due to its inherently strong spin-orbit coupling~\cite{TopologicalInsulator1,TopologicalInsulator2}. To realize the QAHE, the ferromagnetism is prerequisite and can be engineered by magnetic doping~\cite{DMS,proposal4,TI-magnetism1,TI-magnetism2,TI-magnetism3, Qi2016,CodopingEXP}. It was indeed theoretically proposed~\cite{proposal4} in 2010 and later experimentally observed in 2013 in the magnetically doped topological insulator thin films~\cite{ChangCuiZu,ExperimentalQAHE1,ExperimentalQAHE2,2015NatMat,APL}. However, the major obstacle, hindering the practical applications of QAHE, is the extremely low QAHE-observation temperature. Therefore, more efforts are being made to increase the QAHE observation temperature via various doping schemes in topological insulators.

\begin{figure*}
  \includegraphics[width=14.0cm,angle=0]{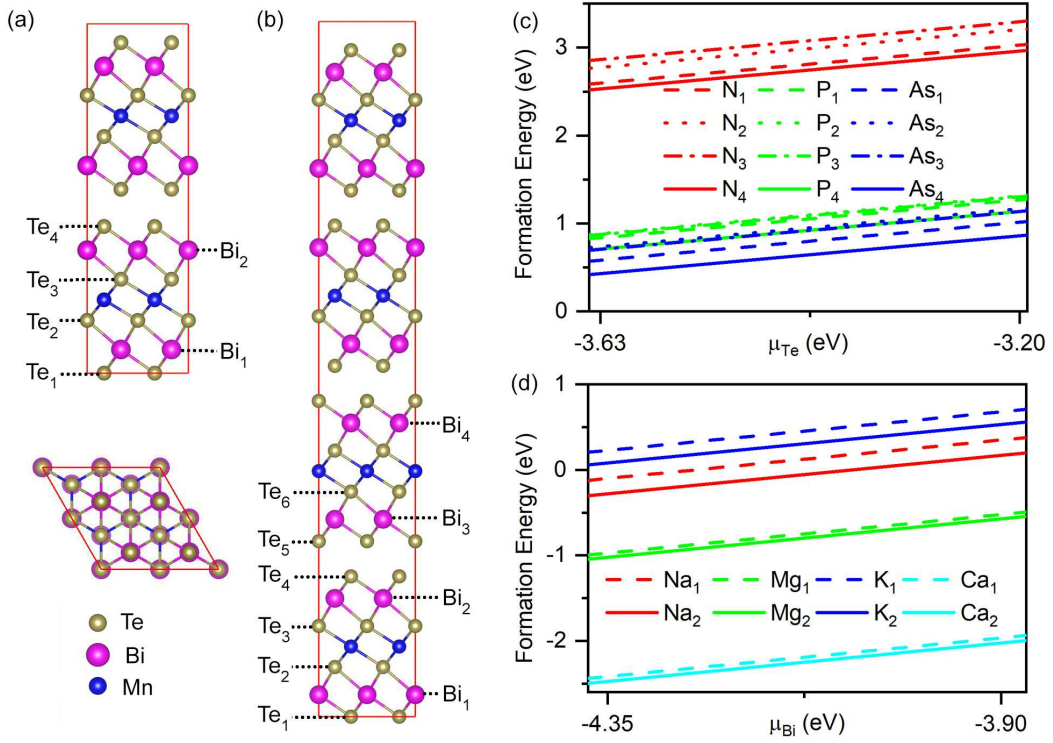}
  \caption{Top view and side views of crystal structures of 2-4 SLs MnBi$_2$Te$_4$ and formation energies of \textit{p}-type doped systems. (a) Two-SL MnBi$_2$Te$_4$ with one N/P/As substitution at sites Te$_1$-Te$_4$, or one Na/Mg/K/Ca substitution at sites Bi$_1$-Bi$_2$; (b) The 3-4 SLs MnBi$_2$Te$_4$ with one N/P/As substitution at sites Te$_1$-Te$_6$, or one Na/Mg/K/Ca substitution at sites Bi$_1$-Bi$_4$. (c)-(d): Formation energies of (c) N/P/As or (d) Na/Mg/K/Ca doped two-SL MnBi$_2$Te$_4$ as a function of the host element chemical potentials.}
\label{Fig1}
\end{figure*}

Alternatively, MnBi$_2$Te$_4$, composed of septuple-layer (SL) blocks stacking along the [0001] direction via van der Waals interaction[see Fig.~\ref{Fig1}(a) and \ref{Fig1}(b)], becomes an appealing host material to realize exotic topological phases~\cite{ZhangHJ-PRL,MBT-HeK,MBT-XuY,ZhangHJ-2}. It exhibits intrinsic magnetism, following the A-type antiferromagnetic order, where the neighboring ferromagnetic Mn layers are coupled in an antiparallel manner~\cite{MBT-HeK,MBT-XuY}. It was reported that the QAHE can be observed at 6.5 Kelvin in a five-SL MnBi$_2$Te$_4$ flake, when an external magnetic field is applied; while the zero-field QAHE can only be observed at 1.4 Kelvin with ultra-high sample quality~\cite{MBT-QAHE,MBT-QAHE-2}. The sensitivity of the QAHE on the sample quality indicates that the interlayer antiferromagnetic coupling is a critical obstacle in the QAHE formation, and the interlayer ferromagnetism is highly desired. The interlayer magnetic coupling of van der Waals materials is determined by the \textit{d}-orbital occupation of transition metals~\cite{doccupation1,doccupation2,doccupation3}. One approach to manipulate the interlayer ferromagnetism is by stacking different \textit{d}-orbital occupied van der Waals materials, e.g., MnBi$_2$Te$_4$/V(Eu)Bi$_2$Te$_4$~\cite{doccupation2,doccupation3}. As demonstrated in below, another most efficient approach is by directly doping nonmagnetic \textit{p}-type elements into MnBi$_2$Te$_4$.

In this Letter, we provide a systematic study on the the magnetic and electronic properties of nonmagnetic \textit{p}-doped MnBi$_2$Te$_4$ multilayers by using first-principles calculation methods. In two-SL MnBi$_2$Te$_4$, the interlayer ferromagnetic coupling can be realized by doping various nonmagnetic \textit{p}-type elements (e.g., N, P, As, and Na, Mg, K, Ca) with the Curie temperature up to $T_{\rm C}=54$ Kelvin. The underlying physical origin is the redistribution of \textit{d}-orbital occupation of Mn element induced hopping channels between $\mathrm{t}_{2g}$ and $\mathrm{e}_g$ from different SLs. Although it is topologically trivial in the \textit{p}-doped two-SL case, the topological phase transition occurs to harbour the high-temperature QAHE with a Chern number of $\mathcal{C}=-1$ when the system thickness is increased, i.e. Ca-doped three-SL, and Ca/Mg-doped four-SL MnBi$_2$Te$_4$, with the interlayer ferromagnetism still being preserved. Our work demonstrates a \textit{p}-doping mechanism in producing ferromagnetism in MnBi$_2$Te$_4$ to form the high-temperature QAHE, which is experimentally accessible.


\begin{figure}
  \includegraphics[width=8.5cm,angle=0]{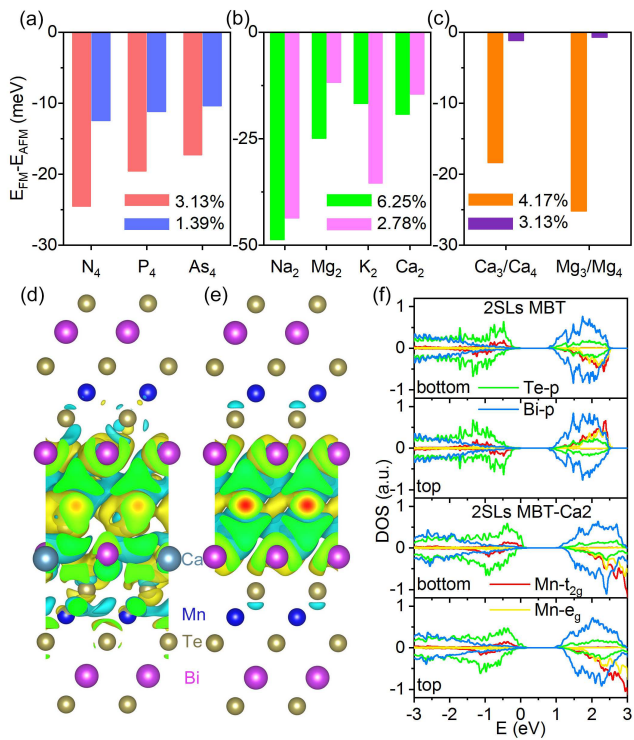}
  \caption{(a)-(c): The energy differences between interlayer ferromagnetic (FM) and interlayer antiferromagnetic (AFM) states of the optimal configurations in (a) N/P/As doped two-SL MnBi$_2$Te$_4$ at 3.13\% and 1.39\% concentrations, (b) Na/Mg/K/Ca doped two-SL MnBi$_2$Te$_4$ at 6.25\% and 2.78\% concentrations, (c) Mg/Ca doped 3-4 SLs MnBi$_2$Te$_4$ at 4.17\% and 3.13\% concentrations. (d)-(e): Differential charge density of (d) Ca doped and (e) pristine two-SL MnBi$_2$Te$_4$. Yellow and green isosurface represent respectively charge accumulation and reduction. (f) Local density of states of Ca doped and pristine two-SL MnBi$_2$Te$_4$. Te-\textit{p}, Bi-\textit{p}, and Mn-\textit{d} orbitals [t$_{2g}$ and e$_{g}$] in each SL of MnBi$_2$Te$_4$ are displayed.
  } \label{Fig2}
\end{figure}

\textit{p-type doping scheme in MnBi$_2$Te$_4$---.} It was known that interlayer magnetic coupling in MnBi$_2$Te$_4$ is dominated by \textit{p}-orbital mediated superexchange interaction, while \textit{d}-orbital occupation has vital influence on the sign of interlayer magnetic coupling~\cite{doccupation2,doccupation3}. Based on the superexchange mechanism, doping \textit{p}-type nonmagnetic elements can change the \textit{d}-orbital occupation of Mn in the same SL. With the aid of hopping channel between 3\textit{d}-orbital of Mn in undoped SL and virtual 3\textit{d}-orbitals of Mn in \textit{p}-doped SL, the interlayer ferromagnetic coupling becomes possible. In experiments, MnBi$_2$Te$_4$ was found to be electron-doping due to their intrinsic \textit{n}-type defects~\cite{MBT-HeK}. Therefore, another natural benefit of \textit{p}-doping is the charge-compensation, which is a prerequisite for realizing the QAHE.

We now first study the possibility of \textit{p}-doping in MnBi$_2$Te$_4$ multilayers. Detail of our first-principles calculations can be found in Supplemental Materials~\cite{SM}. Substituting Te/Bi atoms by nonmagnetic dopants is experimentally feasible, as implemented in Bi$_2$Te$_3$-family topological insulators. The Te and Bi elements in MnBi$_2$Te$_4$ exhibit respectively $2^{-}$ and $3^{+}$ valence states. In order to employ \textit{p}-doping, the corresponding substituted elements can be $3^{-}$, $1^{+}$ and $2^{+}$ valence states, respectively. Therefore, the typical candidates of \textit{p}-type nonmagnetic dopants include N/P/As for Te sites, or Na/Mg/K/Ca for Bi sites. As displayed in Fig.~\ref{Fig1}(a) for a two-SL MnBi$_2$Te$_4$, there are four Te substitutional sites (i.e., Te$_1$, Te$_2$, Te$_3$, Te$_4$), and two Bi substitutional sites (i.e., Bi$_1$, Bi$_2$). The formation energies can be evaluated from the expression~\cite{WGZhu,Nonmagneticdoping}: $\Delta H_{\rm F}=E_{\rm tot}^{\rm D}-E_{\rm tot}-\sum n_{\rm i}\mu_{\rm i}$, where $E_{\rm tot}^{\rm D}$,  $E_{\rm tot}$ are respectively the total energies of the \textit{p}-doped and undoped systems, $\mu_{\rm i}$ is the chemical potential for species $i$ (host atoms or dopants) and $n_{\rm i}$ is the corresponding number of atoms added to or removed
from the system.

Considering the formation energies of N/P/As substitutions at Te sites in one SL as displayed in Fig.~\ref{Fig1}(c), one can find that the Te$_4$ site is preferred. The formation energy of N-substitution (2.5-3.0 eV) is larger than that of either P (0.6-1.2 eV) or As (about 0.4-1.0 eV). For Na/Mg/K/Ca substitutions at Bi sites in Fig.~\ref{Fig1}(d), one can find that the Bi$_2$ site is preferred, and the formation energies of Bi-site substitutions are always lower than those of Te-site substitutions. In particular, the formation energies of Na/Mg/Ca-substituted Bi$_2$ site are negative, suggesting that these dopings are experimentally feasible. As far as we know,
the C-doped ZnO can also be experimentally realized, even thought the estimated formation energy of C substituted O in ZnO is about 5.3 eV~\cite{CZnO}, which is larger than those of aforementioned \textit{p}-type dopants in MnBi$_2$Te$_4$. Hereinbelow, we concentrate on the most stable substitional sites (i.e. Te$_4$ and Bi$_2$) to study the magnetic and electronic properties of \textit{p}-doped two-SL MnBi$_2$Te$_4$.

\textit{Interlayer Ferromagnetism from p-doping---.} Figures~\ref{Fig2}(a) and \ref{Fig2}(b) display the energy differences ($\Delta_E = E_{\mathrm{FM}}-E_{\mathrm{AFM}}$) between interlayer ferromagnetic (FM) and antiferromagnetic (AFM) states of the optimal configurations at different \textit{p}-doped concentrations in two-SL MnBi$_2$Te$_4$. In the absence of doping, the two-SL MnBi$_2$Te$_4$ indeed exhibits interlayer antiferromagnetism~\cite{SM}. The introduction of \textit{p}-type dopants leads to $\Delta_E < 0$, strongly indicating that interlayer ferromagnetic state is more stable than the interlayer anti-ferromagnetic state. For N/P/As doping at Te$_4$ site [see Fig.~\ref{Fig2}(a)], $\Delta_E$ change respectively from -12.4/-11.2/-10.3 meV to -24.5/-19.5/-17.2 meV, along with the increase of doping concentration. For Na/Mg/K/Ca substitution at Bi$_2$ site [see Fig.~\ref{Fig2}(b)], $\Delta_E$ are respectively -43.5/-11.8/-35.4/-14.5 meV at 2.78\% doping concentration, and -48.6/-24.9/16.7/-19.2 meV at 6.25\% doping concentration. Besides the energy difference for optimal doping sites, $\Delta_E$ of all allowed doping sites for different \textit{p}-type elements show that the interlayer ferromagnetic coupling is always preferred [see Fig. S1]~\cite{SM}.

In addition, the ferromagnetic Curie temperature plays a crucial role in determining the QAHE observation temperature. The estimated Curie temperature from mean-field theory is listed in Table S1~\cite{SM}, which ranges between 15.7 and 53.7 Kelvin depending on the dopants. For example, the Curie temperature of Ca-doped MnBi$_2$Te$_4$ can reach $T_{\rm C}=21.2$ Kelvin at 6.25\% doping concentration, which can be further raised with the increase of doping concentration. Note that the higher doping concentration may decrease the spin-orbit coupling of the whole system.

For thicker MnBi$_2$Te$_4$ films (i.e., three-SL and four-SL films), we calculate the $\Delta_E$ of two most favorable dopants (Mg and Ca). For different substitutional sites, it is found that Bi$_3$ (Bi$_4$) site is most stable in three-SL (four-SL) MnBi$_2$Te$_4$ films. And for different magnetic configurations of the most stable doping site, the energy differences show that the ferromagnetic states are preferred [see Tables S2 and S3]~\cite{SM}. Figure~\ref{Fig2}(c) displays the energy difference $\Delta_E$ of one Mg or Ca dopant at Bi$_3$ (Bi$_4$) site in $2\times2$ supercells of three-SL (four-SL) MnBi$_2$Te$_4$. One can see that ferromagnetic coupling strength is dependent on doping concentration that is determined by the number of layers, i.e., for one dopant the increase of septuple layers leads to rapidly decrease of ferromagnetic coupling strength. Therefore, larger ferromagnetic coupling strength in a multilayer system requires higher \textit{p}-type doping concentration.

The formation mechanism of interlayer ferromagnetic coupling can be explained from the differential charge density and local density of states. Let us take the Ca-doped two-SL MnBi$_2$Te$_4$ as an example [see in Fig.~\ref{Fig2}(d)]. In the pristine case [see Fig.~\ref{Fig2}(e)], the charge distribution in the top SL is the same as that in the bottom SL. After Ca-doping in bottom SL, the charge of Mn atoms in the same SL is clearly decreased whereas that in top SL remains nearly unchanged. Such a charge redistribution leads to new hopping channels between Mn atoms in adjacent SLs. In pristine case, the t$_{2g}$ and e$_{g}$ orbitals are fully occupied, leading to the absence of electron hopping between t$_{2g}$ and e$_{g}$ orbitals. While as displayed in Fig.~\ref{Fig2}(f), the decrease of \textit{d}-orbital occupation in bottom SL generates new hopping channels from t$_{2g}$ (top SL) to e$_{g}$ (bottom SL) and e$_{g}$ (top SL) to e$_{g}$ (bottom SL), which are allowed for ferromagnetic coupling. In addition, in Fig.~\ref{Fig2}(f), one can also find that a large spin polarization appears in the Te element after Ca-doping, which suggests that the interlayer ferromagnetic coupling in Ca-doped two-SL MnBi$_2$Te$_4$ is mediated via the interlayer Te-Te superexchange interaction.

\begin{figure}
  \includegraphics[width=8cm,angle=0]{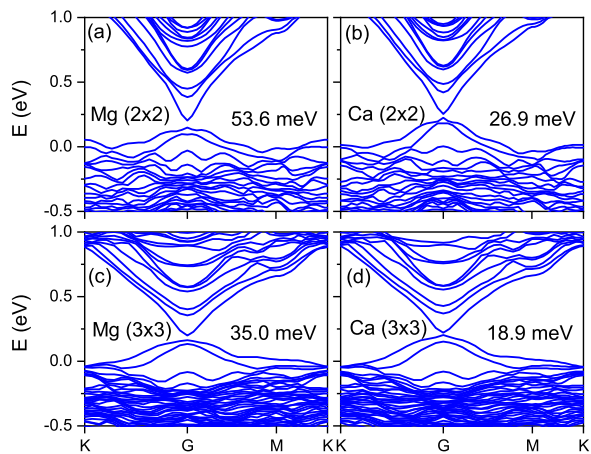}
  \caption{Band structures and corresponding band gaps of Mg- and Ca-doped 2 SLs MnBi$_2$Te$_4$ with optimal configurations along high-symmetry lines. (a)-(b): doping one Mg or Ca atom in $2\times 2$ two-SL MnBi$_2$Te$_4$ with the concentration of 6.25\%. (c)-(d): doping one Mg or Ca atom in $3\times 3$ two-SL MnBi$_2$Te$_4$ with the concentration of 2.78 \%.} \label{Fig3}
\end{figure}

\textit{Electronic structures and topological properties---.} Next, we explore the electronic band structures of the Mg and Ca doped multi-SL MnBi$_2$Te$_4$ [see Supplemental Materials for band structures of other \textit{p}-dopants]~\cite{SM}. Figure~\ref{Fig3} displays the band structure along high-symmetry lines of the optimal configurations of Mg and Ca doped two-SL MnBi$_2$Te$_4$. As illustrated in Figs.~\ref{Fig3}(a) and \ref{Fig3}(b), a band gap about 53.6 meV (26.9 meV) opens at $\Gamma$ point with Mg (Ca) dopant for a doping concentration of 6.25\%. When the doping concentration reduces to 2.78\%, the band gap decreases to about 35.0 meV (18.9 meV) for Mg (Ca) dopant [see Figs.~\ref{Fig3}(c) and \ref{Fig3}(d)]. To verify whether such a gap can host the QAHE or not, one can directly calculate the anomalous Hall conductance $\sigma_{xy}$ by integrating Berry curvature of the occupied valence bands~\cite{BerryCurvature1,BerryCurvature2}. Unfortunately, we obtained $\sigma_{xy}=0 e^2/h$ for all \textit{p}-doped two-SL MnBi$_2$Te$_4$, indicating that it is still a topological trivial phase, even though the ferromagnetism is well established. The possible reasons include: i) the decrease of spin-orbit coupling originated from the light doping elements, and ii) the film thickness influence~\cite{MBT-QAHE-2,MBT-Otrokov-PRL}. To address these concerns, we first choose to dope some heavy metal elements (i.e., Sn, Pb, In, Tl) in two-SL MnBi$_2$Te$_4$ systems~\cite{SM}. It shows that that doping In or Tl results in the interlayer anti-ferromagnetic coupling; whereas although doping Sn or Pb gives rise to interlayer ferromagnetic coupling, no band gap opens at moderate doping concentrations. We then consider the influence of film thickness of MnBi$_2$Te$_4$ in below.

\begin{figure}
  \includegraphics[width=8.5cm,angle=0]{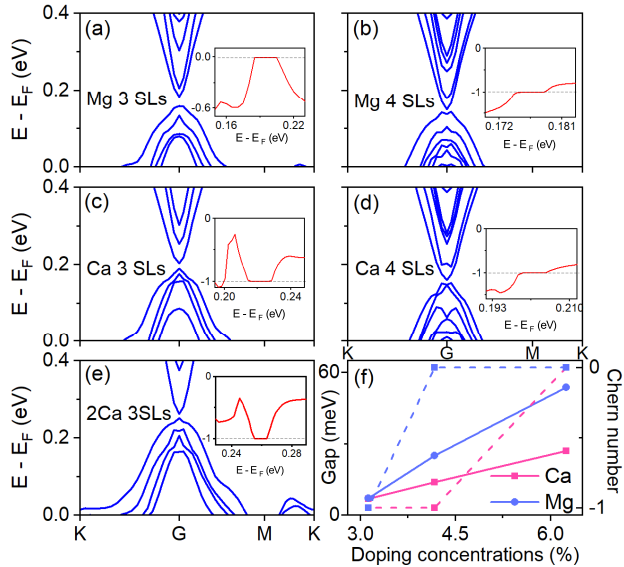}
  \caption{Band structures along high-symmetry lines of  MnBi$_2$Te$_4$ doped with Mg in (a) three-SL and (b) four-SL, doped with Ca in (c) three-SL and (d) four-SL, and (e) doped with two Ca in three-SL. The inset displays the anomalous Hall conductivity as a function of Fermi energy. (f) The dependence of band gap (solid lines) and Chern number (dashed lines) on the doping concentrations of Mg/Ca. The decrease of doping concentration indicates the increase of number of MnBi$_2$Te$_4$ SL.
  } \label{Fig4}
\end{figure}

Figures~\ref{Fig4}(a) and \ref{Fig4}(b) display respectively the band structures of Mg-doped three- and four-SL MnBi$_2$Te$_4$, where the corresponding band gaps are respectively 24.9~meV (at 4.17\% doping concentration) and 7.0~meV (at 3.13\% doping concentration). The Hall conductance $\sigma_{xy}$ evaluation gives respectively 0 and -1 in the units of $e^2/h$ for three-SL and four-SL Mg-doped systems, strongly signaling a topological phase transition from trivial insulator to the QAHE with the increase of film thickness. For the Ca-doped cases as displayed in Figs.~\ref{Fig4}(c) and \ref{Fig4}(d), one can see that the band gaps are respectively 13.7 meV (three-SL) and 6.8 meV (four-SL). Surprisingly, the Hall conductance in the band gap is $\sigma_{xy}=-e^2/h$ for both three- and four-SL Ca-doped MnBi$_2$Te$_4$.
Therefore, the increase of film thickness can lead to a topological phase transition in \textit{p}-doped MnBi$_2$Te$_4$ multilayers.

For three-SL Ca-doped system, we also investigate the role of doping concentration on the electronic properties by including two Ca dopants at different substitutional sites~\cite{SM}. As illustrated in Fig.~\ref{Fig4}(e), the band gap slightly decreases to 11.2 meV, with the nontrivial topology being preserved, but the Curie temperature is greatly enhanced from 20.3 K (one Ca dopant) to 52.0 K (two Ca dopants).
Fig.~\ref{Fig4}(f) summarizes the band gaps and Hall conductance as functions of doping concentration. Compared with Mg dopant, the Ca doped MnBi$_2$Te$_4$ is preferred since that the topological phase appears in system with thinner thickness. Therefore, the Mg- and Ca-doped MnBi$_2$Te$_4$ multilayers are beneficial for realizing the high-temperature QAHE.

\textit{Summary---.}
In conclusion, we propose that a feasible \textit{p}-type doping strategy in MnBi$_2$Te$_4$ can be used to realize interlayer ferromagnetism and the high-temperature QAHE. We provide proof-of-principle numerical demonstration that: (1) interlayer ferromagnetic transition can appear when some nonmagnetic \textit{p}-type elements are doped into MnBi$_2$Te$_4$; (2) band structures and topological property calculations show that Ca- and Mg-doped MnBi$_2$Te$_4$ multilayer can realize the QAHE with Chern number of $\mathcal{C}=-1$.

Experimetally, Mg, Ca and some nonmagnetic elements doped topological insulators have been successfully fabricated in order to tune carrier type and density~\cite{doping1,doping2,doping3}. For example, to compensate the \textit{n}-type carrier induced by Se vacancies in topological insulator Bi$_2$Se$_3$, a small concentration of Ca is doped, and insulating behavior is preserved whereas the Fermi level is tuned into the band gap~\cite{doping1}. For the MnBi$_2$Te$_4$, from our calculation, the formation energies of Ca substitution are only -2.5 to -3.0 eV. Hence the \textit{p}-type Ca dopants in MnBi$_2$Te$_4$ are very feasible in experiment. The merits of \textit{p}-type doping in MnBi$_2$Te$_4$ is that it can not only result in interlayer ferromagnetic coupling without introducing additional magnetic disorder, but also compensate the intrinsic \textit{n}-type carrier, which in principle guarantees the insulating state and is beneficial to realize the high-temperature QAHE in MnBi$_2$Te$_4$. Our work provide a highly desirable scheme to overcome the difficulty of the observing of the QAHE in MnBi$_2$Te$_4$ without applying external magnetic field.

\begin{acknowledgments}
This work was financially supported by the NNSFC (No.11974098, No. 11974327 and No. 12004369), Natural Science Foundation of Hebei Province (A2019205037), China Postdoctoral Science Foundation (2020M681998) and Science Foundation of Hebei Normal University (2019B08), Fundamental Research Funds for the Central Universities, Anhui Initiative in Quantum Information Technologies. The Supercomputing services of AM-HPC and USTC are gratefully acknowledged.
\end{acknowledgments}


\begin{thebibliography}{99}
\bibitem{weng2015quantum}
 H. Weng, R. Yu, X. Hu, X. Dai, and Z. Fang, Adv. Phys. \textbf{64}, 227 (2015).

\bibitem{ren2016topological}
 Y. Ren, Z. Qiao, and Q. Niu, Rep. Prog. Phys. \textbf{79}, 066501 (2016).

\bibitem{he2018topological}
 K. He, Y. Wang, and Q.-K. Xue, Annu. Rev. Condens. Matter Phys. \textbf{9}, 329 (2018).

\bibitem{Haldane}
F. D. M. Haldane, \text{Phys. Rev. Lett.} \textbf{61,} 2015 (1988).

\bibitem{Graphene2004}
K. S. Novoselov, A. K. Geim, S. V. Morozov, D. Jiang, Y. Zhang, S. V. Dubonos, I. V. Grigorieva, A. A. Firsov, Science \textbf{306}, 666 (2004).

\bibitem{proposal1}
 M. Onoda and N. Nagaosa, \text{Phys. Rev. Lett.} \textbf{90,} 206601 (2003).

\bibitem{proposal2}
C. X. Liu, X. L. Qi, X. Dai, Z. Fang, and S. C. Zhang,  \text{Phys. Rev. Lett.} \textbf{101,} 146802 (2008).

\bibitem{proposal3}
C. Wu, \text{Phys. Rev. Lett.} \textbf{101,} 186807 (2008).

\bibitem{proposal4}
 R. Yu, W. Zhang, H. J. Zhang, S. C. Zhang, X. Dai, and Z. Fang, \text{Science} \textbf{329}, 61-64 (2010).

\bibitem{proposal5}
 Z. H. Qiao, S. A. Yang, W. X. Feng, W.-K. Tse, J. Ding, Y. G. Yao, J.Wang, and Q. Niu, \text{Phys. Rev. B} \textbf{82}, 161414(R) (2010).

\bibitem{proposal6}
 Z. F. Wang, Z. Liu, and F. Liu, \text{Phys. Rev. Lett.} \textbf{110,} 196801 (2013).

\bibitem{proposal7}
K. F. Garrity and D. Vanderbilt, \text{Phys. Rev. Lett.} \textbf{110,} 116802 (2013).

\bibitem{proposal8}
 J. Hu, Z. Zhu, and R. Wu,  \text{Nano Lett.} \textbf{15,} 2074 (2015).

\bibitem{proposal9}
C. Fang, M. J. Gilbert, and B. A. Bernevig, \text{Phys. Rev. Lett.} \textbf{112,} 046801 (2014).

\bibitem{proposal10}
 J. Wang, B. Lian, H. Zhang, Y. Xu, and S. C. Zhang, \text{Phys. Rev. Lett.} \textbf{111,} 136801 (2013).

\bibitem{proposal11}
G. F. Zhang, Y. Li, and C. Wu, \text{Phys. Rev. B} \textbf{90,} 075114 (2014).

\bibitem{proposal12}
 H. Z. Lu, A. Zhao, and S. Q. Shen, \text{Phys. Rev. Lett.} \textbf{111,} 146802 (2013).

\bibitem{TopologicalInsulator1}
M. Z. Hasan and C. L. Kane,  \text{Rev. Mod. Phys.} \textbf{82}, 3045-3067 (2010).

\bibitem{TopologicalInsulator2}
 X. L. Qi and S. C. Zhang,  \text{Rev. Mod. Phys.} \textbf{83}, 1057-1110 (2011).

\bibitem{DMS}
T. Jungwirth, J. Sinova, J. Masek, J. Kucera and A. H. MacDonald, \text{Rev. Mod. Phys.} \textbf{78}, 809-863 (2006).

\bibitem{TI-magnetism1}
Y. S. Hor, P. Roushan, H. Beidenkopf, J. Seo, D. Qu, J. G. Checkelsky, L. A. Wray, D. Hsieh, Y. Xia, S. Y.
Xu, D. Qian, M. Z. Hasan, N. P. Ong, A. Yazdani, and R. J. Cava, \text{Phys. Rev. B} \textbf{81}, 195203 (2010).

\bibitem{TI-magnetism2}
C. Niu, Y. Dai, M. Guo, W. Wei, Y. Ma, and B. Huang, \text{Appl. Phys. Lett.} \textbf{98}, 252502 (2011).

\bibitem{TI-magnetism3}
P. P. J. Haazen, J. B. Laloe, T. J. Nummy, H. J. M. Swagten, P. Jarillo-Herrero, D. Heiman, and J.
S. Moodera, \text{Appl. Phys. Lett.} \textbf{100}, 082404 (2012).

\bibitem{Qi2016}
S. Qi, Z. Qiao, X. Deng, E. D. Cubuk, H. Chen, W. Zhu, E. Kaxiras, S. Zhang, X. Xu, and Z. Zhang, Phys. Rev. Lett. \textbf{117}, 056804 (2016).

\bibitem{CodopingEXP}
Y. Ou, C. Liu, G. Y. Jiang, Y. Feng, D. Y. Zhao, W. X. Wu, X. X. Wang, W. Li, C. L. Song, L. L. Wang, W. B.
Wang, W. D. Wu, Y. Y. Wang, K. He, X. C. Ma, and Q. K. Xue, \text{Adv. Mater.} \textbf{30}, 1703062 (2018).

\bibitem{ChangCuiZu}
C.-Z. Chang, J. S. Zhang, X. Feng, J. Shen, Z. C. Zhang, M. Guo, K. Li, Y. Ou, P. Wei, L.-L. Wang, Z.-Q. Ji, Y. Feng, S.
H. Ji, X. Chen, J. F. Jia, X. Dai, Z. Fang, S.-C. Zhang, K. He, Y. Y. Wang, L. Lu, X.-C. Ma, and Q.-K. Xue, \text{Science} \textbf{340}, 167 (2013).

\bibitem{ExperimentalQAHE1}
J. G. Checkelsky, R. Yoshimi, A. Tsukazaki, K. S. Takahashi,
Y. Kozuka, J. Falson, M. Kawasaki, and Y. Tokura, \text{Nat. Phys.} \textbf{10}, 731 (2014).

\bibitem{ExperimentalQAHE2}
X. Kou, S.-T. Guo, Y. Fan, L. Pan, M. Lang, Y. Jiang, Q. Shao, T. Nie, K. Murata, J. Tang, Y. Wang, L. He, T.-K. Lee, W.-L.
Lee, and K. L. Wang, \text{Phys. Rev. Lett.} \textbf{113}, 137201 (2014).

\bibitem{2015NatMat}
 C. Z. Chang, W. Zhao, D. Y. Kim, H. Zhang, B. A. Assaf, D. Heiman, S.-C. Zhang,C. Liu, M. H. W. Chan, and J. S. Moodera,  \text{Nat. Mater.} \textbf{14}, 473 (2015).

\bibitem{APL}
M. Mogi, R. Yoshimi, A. Tsukazaki, K. Yasuda, Y. Kozuka, K. S. Takahashi, and Y. Tokura, Appl. Phys. Lett. \textbf{107}, 182401 (2015).

\bibitem{ZhangHJ-PRL}
 D. Zhang, M. Shi, T. Zhu, D. Xing, H. Zhang, and J. Wang \text{Phys. Rev. Lett.} \textbf{122,} 206401 (2019).

\bibitem{MBT-HeK}
Y. Gong, J. Guo, J. Li, K. Zhu, M. Liao, X. Liu, Q. Zhang, L. Gu, L. Tang, X. Feng, D. Zhang, W. Li, C. Song, L. Wang,
P. Yu, X. Chen, Y. Wang, H. Yao, W. Duan, Y. Xu, S.-C. Zhang, X. Ma, Q.-K. Xue, K. He, \text{Chin. Phys. Lett.} \textbf{36,} 076801 (2019).

\bibitem{MBT-XuY}
J. Li, Y. Li, S. Du, Z. Wang, B. L. Gu, S. -C. Zhang, K. He, W. Duan, Y. Xu, \text{Sci. Adv.} \textbf{5,} eaaw5685 (2019).

\bibitem{ZhangHJ-2}
H. Wang, D. Wang, Z. Yang, M. Shi, J. Ruan, D. Xing, J. Wang, and H. Zhang, \text{arXiv} \textbf{1907,} 03380 (2019).

\bibitem{MBT-QAHE}
Y. Deng, Y. Yu, M. Z. Shi, J. Wang, X. H. Chen, and Y. Zhang, \text{Science} \textbf{367,} 895 (2020).

\bibitem{MBT-QAHE-2}
J. Ge, Y. Liu, J. Li, H. Li, T. Luo, Y. Wu, Y. Xu, and J. Wang, Natl. Sci. Rev. \textbf{7}, 1280 (2020).

\bibitem{doccupation1}
J. W. Xiao and B. H. Yan \text{2D Mater.} \textbf{7,} 045010 (2020).

\bibitem{doccupation2}
Z. Li, J. Li, K. He, X. Wan, W. Duan, and Y. Xu, \text{Phys. Rev. B} \textbf{102}, 081107(R) (2020).

\bibitem{doccupation3}
W. Zhu, C. Song, L. Liao, Z. Zhou, H. Bai, Y. Zhou, and F. Pan, \text{Phys. Rev. B} \textbf{102}, 085111 (2020).

\bibitem{MBT-Otrokov-PRL}
M. M. Otrokov, I. P. Rusinov, M. Blanco-Rey, M. Hoffmann, A. Y. Vyazovskaya, S. V. Eremeev, A. Ernst, P. M. Echenique, E. V. Chulkov, Phys. Rev. Lett. \textbf{122}, 107202 (2019).

\bibitem{WGZhu}
J. M. Zhang, W. G. Zhu, Y. Zhang, D. Xiao, and Y. G. Yao, \text{Phys. Rev. Lett.} \textbf{109}, 266405 (2012).

\bibitem{Nonmagneticdoping}
 S. Qi, R. Gao, M. Chang, T. Hou, Y. Han, and Z. Qiao, \text{Phys. Rev. B} \textbf{102}, 085419 (2020).

\bibitem{CZnO}
 H. Pan, J. B. Yi, L. Shen, R. Q. Wu, J. H. Yang, J.Y. Lin, Y. P. Feng, J. Ding, L. H. Van, and J. H. Yin, \text{Phys. Rev. Lett.} \textbf{99}, 127201 (2007).

\bibitem{BerryCurvature1}
Y. G. Yao, L. Kleinman, A. H. MacDonald, J. Sinova, T. Jungwirth, D.-S. Wang, E. Wang, and Q. Niu, \text{Phys. Rev. Lett.} \textbf{92}, 037204 (2004).

\bibitem{BerryCurvature2}
D. Xiao, M. Chang, and Q. Niu, \text{Rev. Mod. Phys.} \textbf{82}, 1959-2007 (2010).

\bibitem{doping1}
Z. Wang, T. Lin, P. Wei, X. Liu, R. Dumas, K. Liu, and J. Shi, \text{Appl. Phys. Lett.} \textbf{97}, 042112 (2010).

\bibitem{doping2}
S. Byun, J. Cha, C. Zhou, Y. K. Lee, H. Lee, S. H. Park, W. B. Lee, and I. Chung, \text{J. Sol. Sta. Chem.} \textbf{269}, 396 (2019).

\bibitem{doping3}
J. Moon, Z. Huang, W. Wu, and S. oh, \text{Phys. Rev. Mater.} \textbf{4}, 024203 (2020).

\bibitem{SM}
See Supplemental Material

\end{thebibliography}
\end{document}